\documentclass[aps,pre,twocolumn,amsmath,amssymb,superscriptaddress,floatfix]{revtex4}

\usepackage{graphicx}
\usepackage{bm}
\usepackage[usenames]{color}
\bibstyle{apsrev.bib}

\newcommand{\be}{\begin{equation}}
\newcommand{\ee}{\end{equation}}
\newcommand{\beqn}{\begin{eqnarray}}
\newcommand{\eeqn}{\end{eqnarray}}

\begin{document}

\title{Population boundary across an environmental gradient: Effects of quenched disorder}
\author{R\'obert Juh\'asz}
\email{juhasz.robert@wigner.mta.hu}
\affiliation{Wigner Research Centre for Physics, Institute for Solid State Physics and Optics, H-1525 Budapest, P.O.Box 49, Hungary}

\author{Istv\'an A. Kov\'acs}
\email{istvan.kovacs@northwestern.edu}
\affiliation{Department of Physics and Astronomy, Northwestern University, 2145 Sheridan Road, Evanston, IL 60208-3112, USA}
\affiliation{Network Science Institute and Department of Physics, Northeastern University, 177 Huntington Avenue, Boston, MA 02115, USA}
\affiliation{Wigner Research Centre for Physics, Institute for Solid State Physics and Optics, H-1525 Budapest, P.O.Box 49, Hungary}
\affiliation{Department of Network and Data Science, Central European University, N\'ador u. 9, H-1051 Budapest, Hungary}

\date{\today}

\begin{abstract}
Population boundary is a classic indicator of climatic response in ecology. In addition to known challenges, the spatial and dynamical characteristics of the boundary are not only affected by the spatial gradient in the environmental factors, but also by local heterogeneities in the regional characteristics. Here, we capture the effects of quenched heterogeneities on the ecological boundary with the disordered contact process in one and two dimensions with a linear spatial trend in the local control parameter. We apply the strong-disorder renormalization group method to calculate the sites occupied with an $O(1)$ probability in the stationary state, readily yielding the population front's position as the outermost site locally as well as globally for the entire boundary. We show that under a quasistatic change of the global environment, mimicking climate change, the front advances intermittently: long quiescent periods are interrupted by rare but long jumps. The characteristics of this intermittent dynamics are found to obey universal scaling laws in terms of the gradient, conjectured to be related to the correlation-length exponent of the model. 
Our results suggest that current observations might misleadingly show little to no climate response for an extended period of time, concealing the long-term effects of climate change.  
\end{abstract}

\maketitle

\section{Introduction}
\label{sec:intro}

In ecology much effort has recently been devoted to the study of population dynamics in the presence of an environmental gradient, which means that the environmental conditions that affect the reproduction rate or lifetime vary smoothly in space, along a given direction \cite{lennon,holt,hklmt,oikos}. This situation is typically realized for a plant species or vegetation living on a hillside, for which the environmental conditions become more and more favorable or unfavorable with increasing altitude. As a consequence, the density decreases in the unfavorable direction, and the connected patch of colonized area becomes, at sufficiently low densities, fragmented, i.e. it is typically composed of distinct islands. Moving further in the unfavorable direction the local density vanishes. The boundary (also known as the \emph{range margin}) separating the connected and fragmented zones is the subject of intensive investigations \cite{oborny,gastner}, which is also motivated by the phenomenon of global warming. A global change of environmental conditions results in the a shift of the range margin, and this may help to predict the response of a species to climate change or, conversely, it could be used to monitor climate change, see Refs. \cite{mustin,oborny} and references therein.  

Theoretical modeling is based on spatially explicit metapopulation models, the simplest case of which is the contact process \cite{cp}. Here, habitat patches are represented by a two-dimensional grid, the sites of which can be in two states: either colonized or empty. Colonized sites stochastically colonize their neighboring sites, or go extinct with certain rates. 
The spatially homogeneous variant of the model exhibits a continuous phase transition separating a fluctuating active phase from the absorbing one as the control parameter, chosen as the relative rate of colonization vs. extinction, is varied \cite{liggett,md,odor,hhl}. 
In the presence of a weak gradient, i.e. a spatially slowly varying local control parameter, the bulk phase transition of the homogeneous model is transformed into a spatially explicit transition, with a vanishing local density below a critical coordinate (altitude, see Fig. \ref{2dexample}). Due to the close relation with the critical behavior of the homogeneous model, the properties of the gradient model near the critical coordinate obey a scaling theory, which is characterized by critical exponents that are combinations of standard critical exponents of the homogeneous model \cite{sapoval,oborny}.  

In addition to a gradient, in real systems there are also fine-scale heterogeneities in the environmental conditions, stemming from the depth of soil or the local surface relief, etc. Assuming these factors vary slowly in time compared to the dynamics of the model, these can be incorporated as quenched random reproduction or extinction rates. For the standard, gradient-free contact process, such a quenched disorder is known to alter the critical behavior, giving rise to singularities also in an extended region around the critical point \cite{noest}.
According to a real-space renormalization method, also known as strong-disorder renormalization group (SDRG) \cite{im}, the dynamical scaling of the model in one \cite{hiv} and two dimensions \cite{kovacs} is ultra-slow \cite{moreira} and static exponents differ from those of the clean system, related to the infinite-disorder fixed point of the transformation. 
Whether this type of behavior is valid for any weak randomness requires further investigations. 
So far, all critical exponents are found to be universal in the sense that they are independent of the specific distribution of random parameters, confirmed also by  
large scale simulations, even for relatively weak randomness \cite{vd,vfm,vojta_rev}.  
In addition to the altered dynamical scaling and critical exponents, the quenched model has a number of qualitatively different hallmarks from the clean model. 
A key feature is that the disorder polarizes the density, i.e. typically anchors the activity to certain sites where 
the local density (i.e. occupation probability) is $O(1)$ while on the rest of sites it is negligibly small. 
In this work, we will consider the disordered contact process (DCP) with a gradient in the control parameter, and will use the aforementioned renormalization group method to identify the set of sites with $O(1)$ local density, called as colonized cluster (CC). We then study the width of the front of the CC in one and two spatial dimensions in the stationary state. In addition, mimicking climate change, we will study the evolution of the boundary as the global control parameter is slowly (quasistatically) tuned. According to the results, the boundary positions do not shift smoothly but exhibit significant jumps separated by extended quiescent periods. This phenomenon is reminiscent of punctuated equilibrium in evolution biology \cite{eldredge}. Therefore, in monitoring climate change, no response could mean a quiescent period preceding a large jump.
The distribution of jump lengths is found to have a scaling property in terms of the gradient, which is conjectured to involve the static correlation-length exponent of the model.      

The rest of the paper is organized as follows. The model is introduced in section \ref{sec:model}, and its SDRG treatment is reviewed in section \ref{sec:sdrg}. In section \ref{sec:scaling}, general scaling considerations about the width of the front are presented. 
In section \ref{sec:1d}, the distribution and the shift of the front position under a global change of the control parameter are studied in the one-dimensional model within the SDRG approach through a mapping to time-dependent random walks, while, in section \ref{sec:2d}, we address the same questions in the two-dimensional model by applying the SDRG method numerically. Some details of the calculations are presented in the Appendix.  
Finally, the results are summarized in section \ref{sec:summary}. 

\section{The model}
\label{sec:model}
We consider a variant of the contact process \cite{cp,liggett}. The model is defined on lattice sites which can be in two states, either occupied (colonized) or empty (uncolonized). The dynamics is given by a continuous-time Markov process with two kinds of independent transitions. First, occupied sites can colonize the neighboring (empty) sites, second, occupied sites go spontaneously extinct. We study the simultaneous presence of quenched disorder in the transition rates and a constant gradient in one direction, leading to an average local control parameter that varies linearly in space. 
Due to the universality of the model around criticality, only the local ratio of the transition rates matters, irrespectively from the specific details of how the disorder and gradient are implemented. Therefore, as a convenient choice, we implement the gradient in the extinction rates and assign the quenched disorder to the colonization rates. The extinction rates vary linearly with the $x$ coordinate as 
\be 
\mu(x)=p[1-g(x-x_c)],
\label{mux}
\ee
where $g$ is the gradient strength, $p$ is a global control parameter, and $x_c$ denotes the critical coordinate.  The colonization rate between sites $i$ and $j$, $\lambda_{ij}\in (0,1)$, is an independent, identically distributed random variable. 
Due to the gradient, the average local control parameter, which can be defined as usual for the disordered contact process in terms of the average logarithmic parameters \cite{hiv} as   
\be 
\overline{\Delta}(x)=\overline{\ln[\mu(x)/\lambda]}=\ln\mu(x)-\overline{\ln\lambda},
\ee 
varies with the $x$ coordinate. Here, and in the followings, the overbar denotes an average over quenched disorder. As a result, the mean local density $\overline{\rho}(x)$ also varies with $x$, and, above a critical coordinate $x_c$ where $\overline{\Delta}(x)$ hits the critical control parameter of the gradient-free model ($g=0$), $\overline{\Delta}(x_c)=\ln p-\overline{\ln\lambda}\equiv\Delta_c$, the mean density is non-zero, whereas well below $x_c$ it tends rapidly to zero. 
Altogether, in the vicinity of the critical coordinate, $g|x-x_c|\ll 1$, the average control parameter varies linearly in leading order: 
\be
\overline{\Delta}(x)=\overline{\Delta}(x_c) - g(x-x_c) + O\{[g(x-x_c)]^2\}.
\label{deltax}
\ee
The role of the global control parameter $p$ is merely to tune the critical coordinate $x_c$ to a prescribed position (typically to the middle of the sample in numerical calculations). We are interested in the properties of the boundary region for small gradients (i.e. small slopes of the hillside), especially in the scaling behavior in the limit $g\to 0$. 

In the subsequent analytical calculations, we make restrictions neither on the form of disorder distributions nor on the way how quenched disorder and gradient is particularly realized in the transition rates. The only requirement is that the local control parameter is an independent random variable and its average varies in leading order linearly near $x_c$. Note that the latter requirement is generally fulfilled if the average control parameter varies smoothly with the coordinate $x$.

In the numerical calculations, we will consider finite systems with the coordinate $x$ in the range $[1,L]$, and similarly for the coordinate $y$ in two dimensions. For the latter, we consider periodic boundary condition in the $y$ direction. We use the extinction rate profile given Eq. (\ref{mux}), and fix the gradient to be inversely proportional to the system size: $g=a/L$.
Due to this, the gradient is controlled by the system size, and we can apply standard finite-size scaling techniques. 
With the choice $a=2$, the extinction rate at the edge $x=L$ of the lattice is zero, $\mu(L)=0$, which is also known as active wall boundary condition \cite{hhl}. This choice ensures for the finite systems to have a true non-trivial steady state.  

\section{Determining the colonized cluster by renormalization}
\label{sec:sdrg}

\begin{figure}[ht!]
\includegraphics[width=8.5cm]{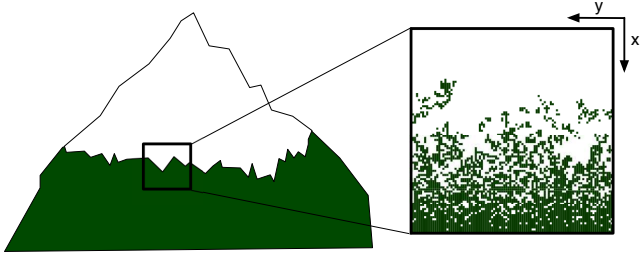}
%Fig_0_small.eps
\caption{\label{2dexample}
We study the boundary of colonized sites across an environmental gradient. The mountain tree line is a highly visible and well studied example (left). In our work, a finite region around the boundary is considered and the spatial and dynamical characteristics of the population boundary is investigated. On the right, we show the results of the contact process simulation in the presence of quenched disorder at size $L=100$.   
}
\end{figure}

A basic question in ecology is to precisely define the range margin of a species or vegetation type exposed to an environmental gradient. 
On the modeling side, one usually samples a random configuration in the stationary state, where some of the sites are occupied while the rest is not (Fig. \ref{2dexample}). Then, treating the sample as a percolation problem, the hull of the percolating cluster can serve as the boundary between the two zones \cite{gastner}. 
The boundary defined this way is a stochastic object which fluctuates in time along with the configuration of the system. It is important to note that the mean $x$ coordinate of the hull, $x_h$ is different from $x_c$, as it is determined by the percolation threshold of typical configurations, which is different from the critical point \cite{gastner}.  

%%%%%%%%%%%%%%%%%%%%%%%%%%%%%%%%%%%%%%%%%%%%%%%%%%%%%%%%%%%%%%%%%%%%%%%%%
%%%%%%%%%%%%%%%%%%%%%%%%%%%%%%%%%%%%%%%%%%%%%%%%%%%%%%%%%%%%%%%%%%%%%%%%

In presence of quenched disorder, the local densities, $\rho_i$, are site dependent. This enables an alternative classification of the sites as those having a local density larger than an arbitrary threshold $\rho_c$ and those having $\rho_i<\rho_c$. The so defined colonized sites have two important differences from the connected zone delineated by the hull. First, these zones do not fluctuate stochastically and are determined by the underlying random environment (and $\rho_c$). Second, since the local densities do not vary monotonically with the coordinates, the active zone is no longer necessarily connected in the percolation sense.   

There is a well-established coarse-graining procedure of the disordered contact process to find the sites which are colonized with a high probability, without requiring to set an arbitrary threshold $\rho_c$.
The method is known as strong-disorder renormalization group in the physics literature \cite{im} and was originally designed for studying low-energy properties of disordered quantum spin chains. Its applicability and power to describe the long-time behavior of the disordered contact process was recognized later \cite{hiv}. 
According to this method, the critical behavior of the model in one and two dimensions, at least for not too weak disorder, is governed by a so called infinite-disorder fixed point of the renormalization transformation, at which the procedure becomes asymptotically exact \cite{im}. 

The renormalization procedure consists of two kinds of steps, which are applied sequentially, depending on whether the momentarily largest parameter of the model is a colonization or an extinction rate. When the maximal rate is a colonization rate ($\lambda_{ij}$), the two sites form a highly correlated cluster, with an effective extinction rate obtained by perturbation calculation, since all other rates are smaller:
\be 
\ln \tilde \mu = \ln \mu_i + \ln\mu_j - \ln\lambda_{ij} + \ln 2. 
\label{mu_rule}
\ee
On the other side, when the maximal rate is an extinction rate $\mu_i$, the site is most of the time extinct, therefore we can remove it from the system, apart from weak induced interactions between its neighboring sites, obtained perturbatively as
\be 
\ln \tilde \lambda_{jk} = \ln \lambda_{ij} + \ln\lambda_{ik} - \ln\mu_i. 
\ee
At the critical point, the logarithmic rates increase in magnitude without limits during the renormalization procedure, therefore the term $\ln 2$ in Eq. (\ref{mu_rule}) will not influence the asymptotic properties and can be safely omitted \cite{sm}. At criticality, the distribution of the logarithmic rates also gets broader, increasing the accuracy of the perturbative approach as the process is performed.
Beyond the one dimensional geometry, the above steps can generate a weak colonization rate between existing sites that have been already connected by a colonization rate. 
To resolve this problem we follow the standard maximum rule, according to which the smaller rate is omitted. An advantage of this approximation is that it is self-consistent and enables a very efficient numerical implementation \cite{kovacs}.  
Then, in both steps the generated effective rates are smaller than the eliminated ones. 

Applying the SDRG to a finite system, some of the sites will be eliminated during the procedure (i.e. rendered inactive), and some survive as constituents of the last remaining cluster the procedure ends up with. 
The sites of this cluster are precisely those which are colonized in the stationary state with an $O(1)$ probability, so they constitute the colonized cluster, while the local density on other (eliminated) sites becomes negligibly small. The active wall boundary condition $\mu(L)=0$, obtained with $a=2$, means that the CC contains the sites at $x=L$, as they are never eliminated. 

The validity of the SDRG method in describing the criticality of the disordered contact process has been confirmed by Monte Carlo simulations in several works, see e.g. \cite{vd,vfm,im}. 
In addition, a direct comparison of the density profile in the stationary state shows a good overlap with the CC of the SDRG in individual samples \cite{kj}.   

\section{Scaling of the width of the front}
\label{sec:scaling}

In particle systems subject to a gradient, such as diffusion \cite{sapoval} or contact process \cite{gastner}, the hull of the percolating cluster fluctuates in time, and the amplitude of deviations in the gradient direction increases as the gradient decreases. Concerning the boundary of the colonized cluster in the disordered contact process, it does not have temporal fluctuations. Yet, it still has sample-to sample fluctuations, and in two dimensions, even in a fixed sample, the $x$ coordinate of the boundary varies along the $y$ direction. Therefore, we can define the width of the boundary even in quenched disordered models. 

The scaling of the width of the diffusion frontier with a gradient was derived in Ref. \cite{sapoval}. The argument, which will be briefly recapitulated, is, however, quite general and applies also to the boundary of the CC in the disordered contact process as follows. Away from the critical coordinate, in the unfavorable zone, the activity is typically restricted to distinct ``islands'', while in the favorable zone, empty patches (``lakes'') appear in the occupied background (Fig. \ref{2dexample}). The characteristic linear size of islands and lakes is given by the local correlation length, $\xi[\overline{\Delta}(x)]$, which varies with $x$ through the local control parameter: 
\be
\xi[\overline{\Delta}(x)]\sim |\overline{\Delta}(x)-\Delta_c|^{-\nu_{\perp}}\sim |g(x-x_c)|^{-\nu_{\perp}},
\label{xi}
\ee 
where $\nu_{\perp}$ is the correlation-length exponent of the model. 
The ruggedness of the population ``shoreline'' can be viewed as a result of large islands in the unfavorable zone reaching and joining to the occupied background and, similarly, large lakes in the favorable zone getting connected with the ``sea''. 
The farthest position from $x_c$ at which this can happen is given by the condition that the characteristic size of islands and lakes at that position is comparable with the distance $|x-x_c|$. This gives the characteristic width of the margin and also the maximal distance, $\ell_x$ as
$\ell_x\sim (g\ell_x)^{-\nu_{\perp}}$,
yielding
\be 
\ell_x\sim g^{-\alpha},
\label{ellg}
\ee
where the exponent $\alpha$ is given by 
\be 
\alpha=\frac{\nu_{\perp}}{1+\nu_{\perp}}.
\ee
The above reasoning holds also for the disordered contact process, for which the characteristic size of islands and lakes is determined by the fluctuations of disorder through the correlation-length exponent of the disordered model. The numerical values of the exponents $\nu_{\perp}$ and $\alpha$ for the clean and disordered CP in one and two dimensions are summarized in Table \ref{table_exponents}. 
%%%%%%%%%%%%%%%%%%%%%%%%%%%%%%%%%%%%%%%%%%%%%%%%%%%%%%%%%%%%%%%%%%%
\begin{table}[ht!]
\begin{center}
\begin{tabular}{|c|l|l|l|l|}
\hline    & 1d CP & 1d DCP & 2d CP & 2d DCP \\
\hline
\hline  $\nu_{\perp}$   &  $1.096854(4)$\cite{jensen} &  $2$ \cite{hiv} & $0.733(4)$ \cite{gz,voigt}  & $1.24(2)$ \cite{kovacs}  \\
\hline  $\alpha$      &    $0.523095(1)$  & $2/3$  & $0.423(1)$  & $0.554(4)$ \\
\hline
\end{tabular}
\end{center}
\caption{\label{table_exponents} Critical exponents $\nu_{\perp}$ and $\alpha$ for the clean (CP) and disordered contact process (DCP) in one and two dimensions.}
\end{table}

The characteristic length $\ell_x$ gives the width of the fluctuations of the front in the $x$ direction, scaling as in Eq. (\ref{ellg}). Besides, it can also be interpreted as a correlation length characterizing the spatial correlations in the gradient direction between $x_c$ and another $x$, across a medium where the local control parameter is continuously changing in space.
This correlation function or the distribution of the $x$ coordinate of the front, as it is derived in section \ref{app:distribution}, has a compressed exponential tail: 
\be 
P(l)\sim e^{-\mathrm{const}\cdot g^{\nu_{\perp}}l^{1+\nu_{\perp}}}\sim e^{-\mathrm{const}\cdot(l/\ell_x)^{1+\nu_{\perp}}}.
\label{Pl}
\ee

The gradient thus gives rise to a finite correlation length $\ell_x$ in the gradient direction. We mention that this will also induce a finite correlation length $\ell_y$ in the $y$ direction at $x=x_c$, which, due to the intrinsic isotropy of the (gradient-free) model, is expected to scale with $g$ in the same way as $\ell_x$:
\be 
\ell_y\sim\ell_x\sim g^{-\alpha}.
\label{elly}
\ee

%%%%%%%%%%%%%%%%%%%%%%%%%%%%%%%%%%%%%%%%%%%%%%%%%%%%%%%%%%%%%%%%%%%%%%%%%%
\section{Disordered contact process in one dimension}
\label{sec:1d}

\subsection{Width of the boundary}
\label{subsec:1dwidth}

We will show that, for the DCP in one dimension, the general result in Eq. (\ref{ellg}) obtained by scaling considerations can be confirmed by a direct calculation within the frames of the SDRG. 
In this case, we will consider the coordinate $x_f$ of the outermost site of the CC, and define the width of the front by the standard
deviation of $x_f$, 
$\ell_x=\sqrt{\overline{x_f^2}-\overline{x_f}^2}$,  
over different samples of disorder realizations.
Again, the front position $x_f$ in a given realization of the random environment is sharp {\it per definitionem}, and we consider here its variation with the underlying random environment.   

\begin{figure}[ht!]
\begin{center}
\includegraphics[width=8cm]{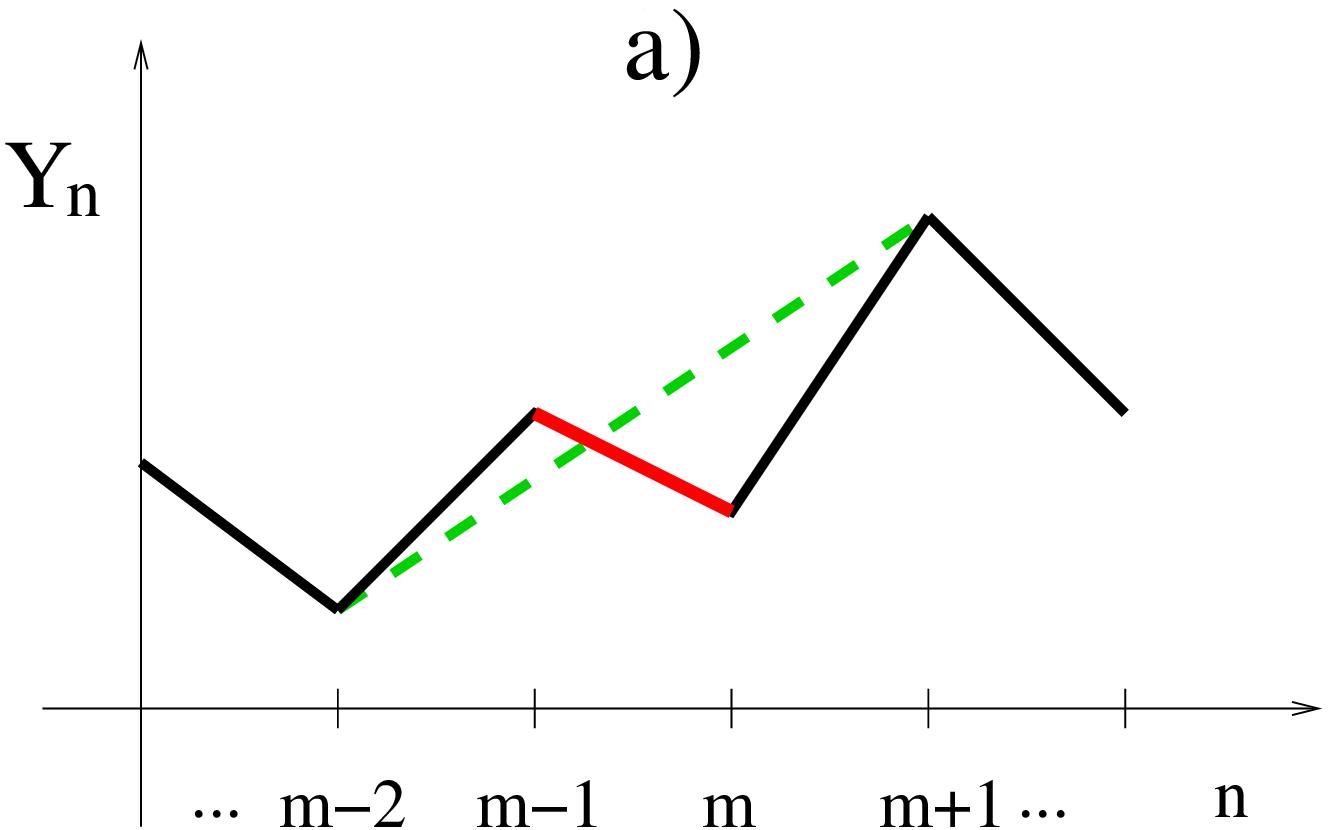}
\includegraphics[width=8cm]{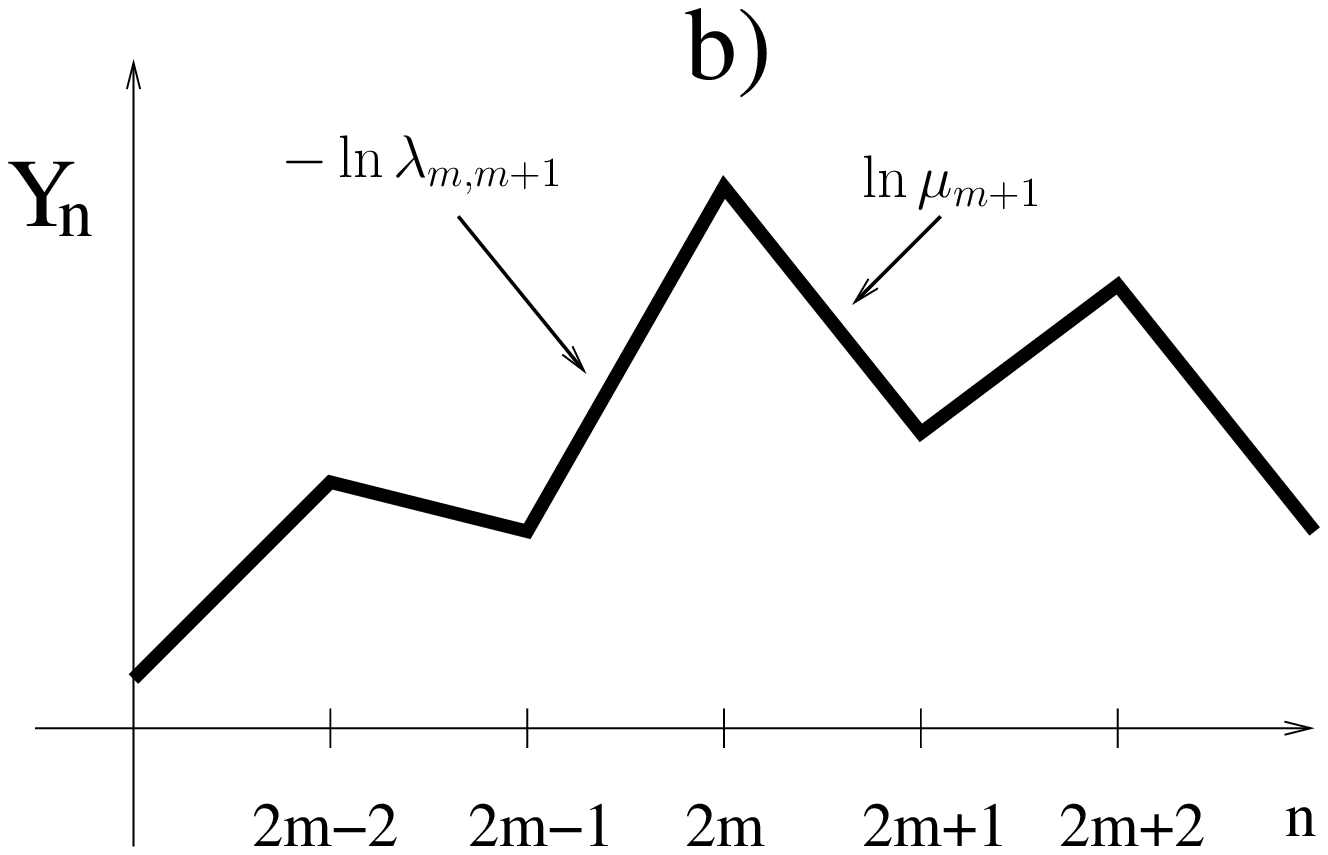} 
%zigzagrg.eps
%maxproof.eps
\caption{\label{zigzag} a) Illustration of an SDRG step by the path $Y_n$, where descending (ascending) segments correspond to extinction (colonization) rates.
The segment of shortest height is shown in red, and, as a result of the SDRG step, the section of the path between $m-2$ and $m+1$ is replaced by a single segment (dashed green line).      
b) Part of the path $Y_n$ around the global maximum at $n=2m$. The segment between $2m$ and $2m+1$ has a height $|\ln\mu_{m+1}|$, while the segment between $2m-1$ and $2m$ has a height $|\ln\lambda_{m,m+1}|$.  
}
\end{center}
\end{figure}

We start with a zig-zag path mapping of the random environment, see e. g. Ref. \cite{jkri} for a similar model. 
Introducing the logarithmic variables 
$\beta_{2x-1}=\ln \mu_{x}$ and $\beta_{2x}=-\ln \lambda_{x,x+1}$,
we define a sequence as 
\be 
Y_n=\sum_{m=n_0}^n\beta_m,
\ee
which is a sum of random terms with alternating signs (Fig. \ref{zigzag}). Here, we implicitly assumed that all rates lie in the interval $(0,1)$. Restricting ourselves to even values of the indices, $n=2x$, the alternation is eliminated, and  $Y_{2x}$ can be regarded as a special, one-dimensional random walk in discrete time $2x$. 
Note that the space variable $x$ of the original problem appears here as the time variable of the random walk. 
The action of the SDRG transformation is particularly simple in terms of the sequence $Y_n$. The shortest in magnitude step of the walk $|\beta_m|=\min\{|\beta_x|\}$ is picked and eliminated together with its two neighbors and replaced by a single step $\tilde\beta=\beta_{m-1}+\beta_m+\beta_{m+1}$ of longer time. 
This corresponds to the coarse-graining of the path $Y_n$ as illustrated in Fig. \ref{zigzag}a. 
%%%%%%%%%%%%%%%%%%%%%%%%%%%%%%%%%%%%%%%%%%%%%%%%%%%%%%%%%%%%%%%%%%%%%%%%%
%%%%%%%%%%%%%%%%%%%%%%%%%%%%%%%%%%%%%%%%%%%%%%%%%%%%%%%%%%%%%%%%%%%%%%%%
The local order parameter is given by $\Delta_x=\ln(\mu_x/\lambda_{x,x+1})=\beta_{2x-1}+\beta_{2x}$.
Within our SDRG approach of the problem, the critical point is given by $\overline{\Delta}=0$, so the critical coordinate (or time) is simply $x_c=0$.  
Thus, in line with Eq. (\ref{deltax}) we have
\be 
\overline{\Delta_x}=-gx
\label{bias}
\ee
at least for $|gx|\ll 1$. 
We assume that the left boundary $x_0=(n_0+1)/2$ of the system, as well as the right one is far from the origin compared to the yet unspecified width $\ell_x$ of the critical zone, so that the system can be regarded as practically infinite. 
At even values of the indices, $n=2x$, we have 
$Y_{2x}=\sum_{m=n_0}^{2x}\beta_m=\sum_{m=x_0}^x\Delta_m$.
This means that the random walk $Y_{2x}$ has a time-dependent bias given in Eq. (\ref{bias}), which drives the walker to the positive (negative) direction for $x<0$ ($x>0$) (Fig. \ref{zigzag}). 
The average path as the function of the discrete time is thus parabolic close to the critical time $x=0$ (Fig. \ref{px}): 
\be 
\overline{Y_{2x}}=-\frac{g}{2}x^2+O(|gx|) + \mathrm{const}.
\label{Yav}
\ee

Now we make the following statement. The time $2m$ at which the path reaches its maximum,  $Y_{2m}=\max_n\{Y_n\}$, determines the position of the front as $x_f=m+1$.
The proof of this statement can be found in \ref{app:rw}. 
According to the approximate calculations based on properties of random walks in \ref{app:rw},  
$P(x_f)$ scales as
\be 
P(x_f)\sim  e^{-const\cdot g^2x_f^3}.
\label{Pxf}
\ee
This means that the width $\ell_x$ of the distribution of $x_f$
scales with $g$ according to 
\be 
\ell_x\sim g^{-2/3}.
\label{xfg}
\ee
These results are consistent with the general scaling results in Eqs. (\ref{ellg}) and (\ref{Pl}), using the correlation-length exponent $\nu_{\perp}=2$ of the DCP. Note that the coordinate $x$ of the original problem corresponds to the time in the equivalent random walk picture, therefore $\nu_{\perp}$ is given by the temporal correlation-length exponent of random walk, see Eq. (\ref{tau}).

At the characteristic distance $\ell_x$ from the critical coordinate, the local control parameter is $|\overline{\Delta_{\ell_x}}|\sim g\ell_x\sim g^{1/3}$,
which tends to zero in the limit $g\to 0$. This further justifies the approximation made in the SDRG rules of the contact process, i.e. the neglection of the term $\ln 2$ in Eq. (\ref{mu_rule}). 

\begin{figure}[ht!]
\includegraphics[width=8.5cm]{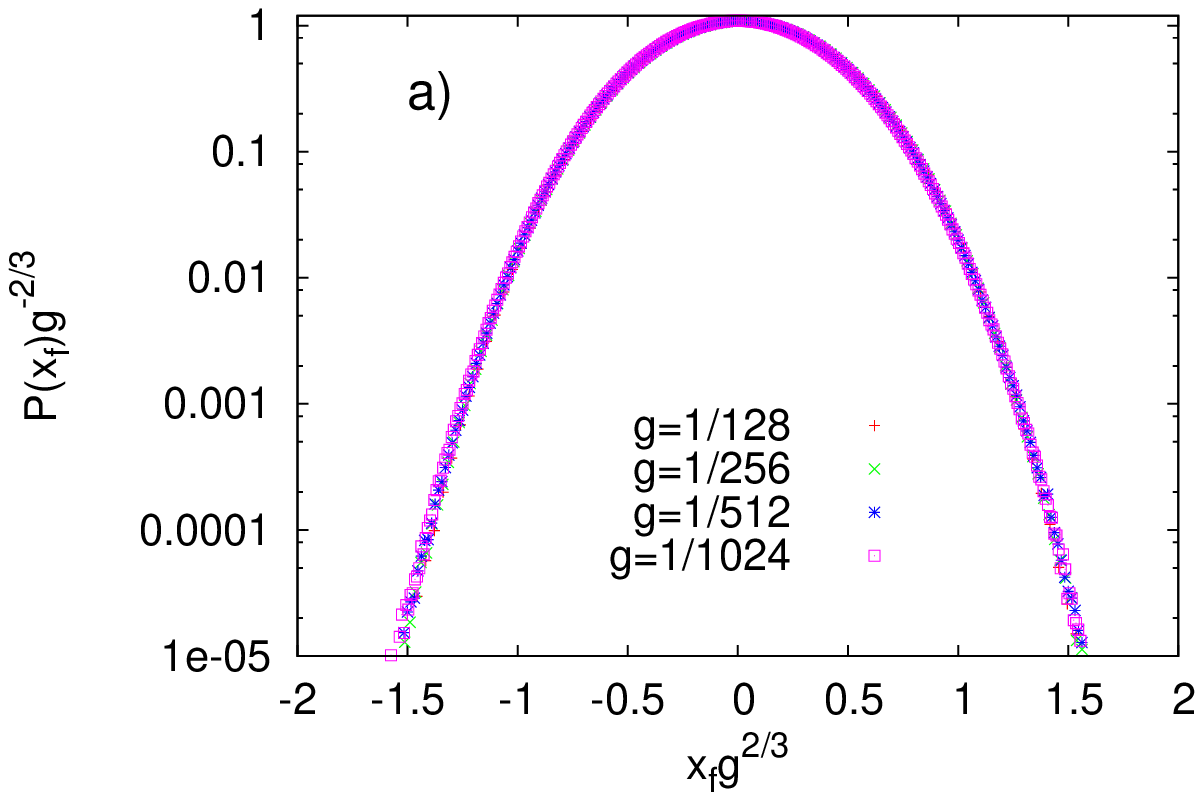} 
\includegraphics[width=8.5cm]{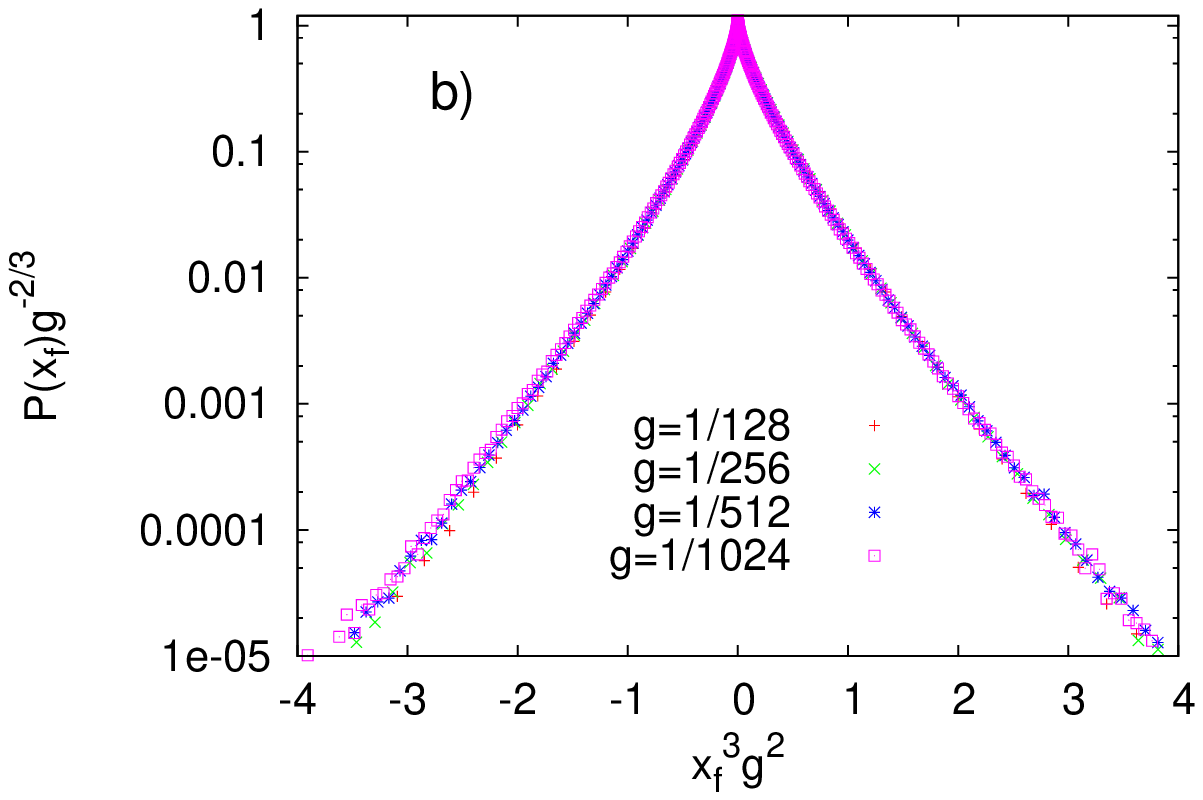} 
%md.eps
%md2.eps
\caption{\label{xfdist} a) The distribution of the scaling variable $x_fg^{2/3}$ obtained by simulations of the time-dependent random walk described in the text. The number of samples was $10^8$ for each value of the gradient.
b) The same data plotted against $x_f^3g^2$. In this plot, the tail of the distribution must be linear according to Eq. (\ref{Pxf}).
}
\end{figure}

To test the assumptions behind Eq. (\ref{Pxf}), we
performed numerical simulations of random walks with a time-dependent bias. 
We have drawn $\Delta_n$ from uniform distributions of unit width with an offset of the support increasing linearly in time, so that the average control parameter changes in time as  $\overline{\Delta_x}=gx$. Distributions of $x_f$ for different values of the gradient $g$ are shown in Fig. \ref{xfdist}. 
The data show a good collapse in terms of the scaling variable $x_fg^{2/3}$. Furthermore, as it is demonstrated in Fig. \ref{xfdist}b, the tail of the distribution is in accordance with the compressed exponential form in Eq. (\ref{Pxf}).
%%%%%%%%%%%%%%%%%%%%%%%%%%%%%%%%%%%%%%%%%%%%%%%%%%%%%%%%%%%%%%%%%%%%%%%%%
%%%%%%%%%%%%%%%%%%%%%%%%%%%%%%%%%%%%%%%%%%%%%%%%%%%%%%%%%%%%%%%%%%%%%%%%

\subsection{Evolution of the front under a global change of the environment}

Motivated by the potential role of range margins in monitoring climate change mentioned in the Introduction, in the following we will examine the question how the front $x_f$ shifts in a fixed (large) random environment when the control parameter is globally changed.
In particular, we examine the effects of quenched disorder in such a scenario. 
To start, we assume a constant average gradient, as before,  
\be 
\overline{\Delta_x}=-g(x-x_c),
\ee
but in a random environment with fixed colonization rates. Then, we tune the critical coordinate $x_c$ by shifting the extinction rate profile in Eq. (\ref{mux}). 
The front position $x_f$ is calculated for a range of $x_c$, in the corresponding stationary states for each value of $x_c$ by the SDRG method, and   
we are interested in the change of $x_f$ under a unit increase in $x_c$. 
This approach corresponds to the quasistatic change of the global environment, i.e. sufficiently slow that the system is able to relax to the instantaneous stationary state.
The validity of this approach is quantitatively discussed in Appendix \ref{app:qs}.

\begin{figure*}
\includegraphics[width=18cm]{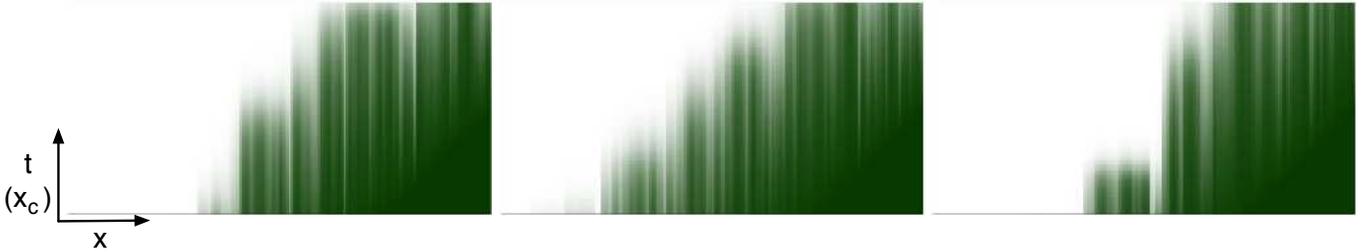}
%profile_small.eps
\caption{\label{sim1d} Snapshots of stationary density profiles obtained by numerical simulations of the DCP in three different samples of size $L=200$, and for a set of $x_c$ in the range $[51,150]$. The horizontal axis corresponds to space, $x=1,\dots,200$, and the density profiles for $x_c=51,\dots,150$ are plotted along the vertical axis.    
}
\end{figure*}

First, to gain a general impression of the change of the density profile under shifting $x_c$, we performed numerical simulations of the DCP in samples of size $L=200$, when $x_c$ is swept from $x_c=51$ to $x_c=150$ in unit steps, leaving sufficient time for the system to relax and for a measurement of the density profile for each value of $x_c$. Density profiles obtained in three different samples are shown in Fig. \ref{sim1d}. 
%\begin{widetext}
%%%%%%%%%%%%%%%%%%%%%%%%%%%%%%%%%%%%%%%%%%%%%%%%%%%%%%%%%%%%%%%%%%%%%%%%%
%%%%%%%%%%%%%%%%%%%%%%%%%%%%%%%%%%%%%%%%%%%%%%%%%%%%%%%%%%%%%%%%%%%%%%%%
%\end{widetext}
In comparison, the SDRG approach classifies the lattice sites to be either practically empty $\rho_i\approx 0$ or fully occupied $\rho_i\approx 1$, with the binary assumption being valid for asymptotically large scales, while the local densities are blurred on a microscopic scale in the simulations. Yet, with increasing system size, the profiles, when viewed on a macroscopic scale, are expected to become more and more contrasted and to approach to the outcome of the SDRG method. 
In spite of these difficulties in small systems, we can clearly observe that for increasing $x_c$, there are idle periods in which the front of the high-density region practically does not shift (just a weak overall decrease of the density occurs), and then abruptly a cluster of sites fades away, resulting in a considerable advance of the front.

\begin{figure}[ht!]
\includegraphics[width=8.5cm]{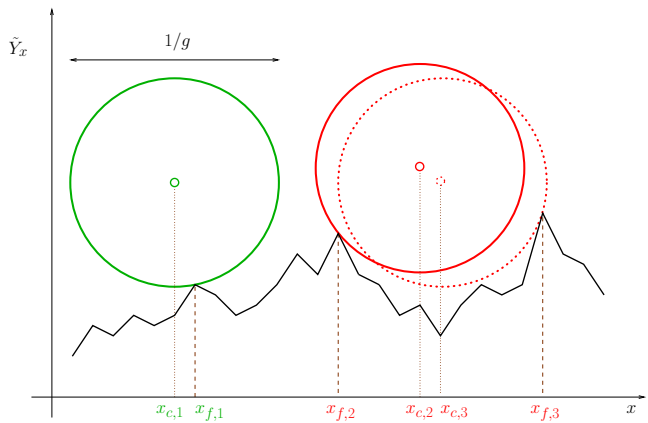} 
%wheel.eps
\caption{\label{wheel} Rolling wheel on a rough surface: an analogous formulation of the problem of determining the front position in the one dimensional DCP within the SDRG approach. The picture shows three different positions of the wheel. The center corresponds to the critical coordinate $x_c$ while the actual touching point with the unbiased path $\tilde Y_x$ gives the front position. As can be seen, changing of the coordinate of the center from $x_{c,2}$ to $x_{c,3}$, the touching point performs a large jump.}
\end{figure}

\begin{figure}[ht]
\includegraphics[width=8.5cm]{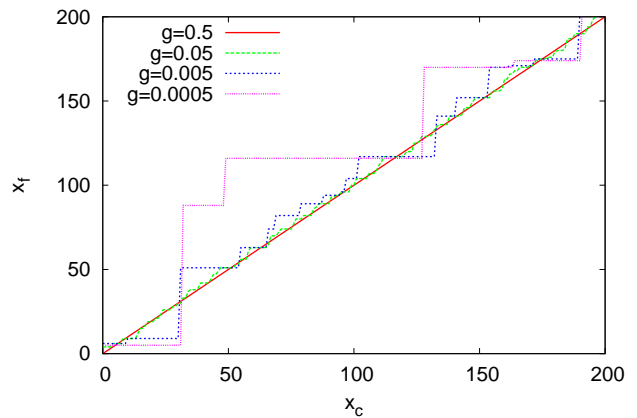} 
%spl1.eps
\caption{\label{spl1} The front position $x_f$ as a function of the critical coordinate $x_c$ determined numerically for different values of the gradient in a fixed random environment. 
}
\end{figure}

\begin{figure}[ht!]
\includegraphics[width=8.5cm]{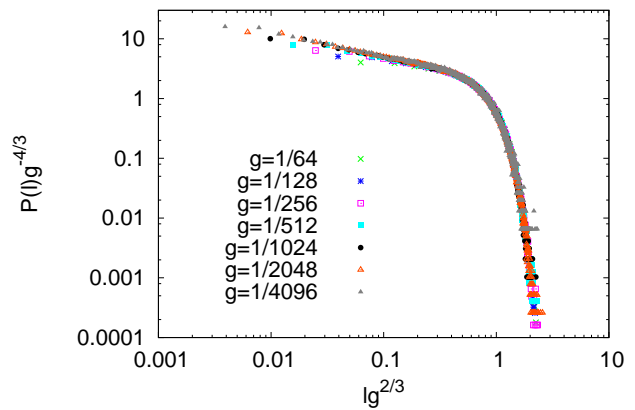} 
%step1.eps
\caption{\label{step1} Scaled distributions of the shift of the front position obtained numerically in the one-dimensional model for different values of the gradient.
}
\end{figure}

Let us continue now with the SDRG treatment of the model. We have seen in the previous section that, for a given value of $x_c$, $x_f$ is determined by the maximum point of the path $Y_x$, or, in other words, by the coordinate at which the path $Y_x$ is touched from above by a straight, horizontal line. The path has a parabolic trend in leading order around $x_c$, $\overline{Y_{x}}=-\frac{g}{2}(x-x_c)^2$.
To get a clearer picture of the problem, this trend can be transferred from the path to the touching line. 
Considering an unbiased random walk path, $\tilde Y_{x}\equiv Y_{x}-\overline{Y_{x}}$, for which $\overline{\tilde Y_x}=0$ for all $x$, $x_f$ is equivalently determined by the touching point of $\tilde Y_{x}$ with a parabola $Y(x)=\frac{g}{2}(x-x_c)^2$.
Going further, for small $g$, the parabola is not much different from a circle of diameter $d=1/g$ in the vicinity of its bottom point, so we may think of the following picture, illustrated in Fig. \ref{wheel}.  
%%%%%%%%%%%%%%%%%%%%%%%%%%%%%%%%%%%%%%%%%%%%%%%%%%%%%%%%%%%%%%%%%%%%%%%%%
%%%%%%%%%%%%%%%%%%%%%%%%%%%%%%%%%%%%%%%%%%%%%%%%%%%%%%%%%%%%%%%%%%%%%%%%
We have a wheel of diameter $d=1/g$, the horizontal coordinate of its center is $x_c$, and it is rolling on a rough (scale-free) 'surface' formed by the unbiased random walk path $\tilde Y_{x}$. The horizontal coordinate of the contact point gives $x_f$. It is now intuitive that the contact point will change intermittently, i.e. for most steps of $x_c$ it remains pinned, while there are certain rare steps at which it suddenly makes a large jump. 
This is also supported by Fig. \ref{spl1}, in which $x_f$ is plotted as a function of $x_c$ for different values of the gradient. 
%%%%%%%%%%%%%%%%%%%%%%%%%%%%%%%%%%%%%%%%%%%%%%%%%%%%%%%%%%%%%%%%%%%%%%%%%
%%%%%%%%%%%%%%%%%%%%%%%%%%%%%%%%%%%%%%%%%%%%%%%%%%%%%%%%%%%%%%%%%%%%%%%%
To obtain quantitative results, $x_f$ was calculated numerically in long samples under unit changes of $x_c$ ($10^7$ steps for each value of $g$). We denote the change in $x_f$ by $l$. According to the results, the 
probability that a non-zero jump occurs scales as
\be
{\rm Prob}(l>0)\sim g^{\kappa},
\label{prob+}
\ee
while the probability that the front remains pinned under a unit change of $x_c$ is
\be
{\rm Prob}(l=0)=1-O(g^{\kappa}).
\label{prob0}
\ee
This means that the shift of the front is a rare event for small gradients, but once it happens, the front makes a long jump of typical length $O(g^{-\kappa})$.
Accordingly, the distribution of jump lengths has the scaling property
\be 
P(l,g)=g^{2\kappa}\tilde P(lg^{\kappa}),
\label{Plg} 
\ee
see Fig. \ref{step1}. The exponent $\kappa$ is found by the scaling collapse to be $\sim2/3$.   
%%%%%%%%%%%%%%%%%%%%%%%%%%%%%%%%%%%%%%%%%%%%%%%%%%%%%%%%%%%%%%%%%%%%%%
%%%%%%%%%%%%%%%%%%%%%%%%%%%%%%%%%%%%%%%%%%%%%%%%%%%%%%%%%%%%%%%%%%%%%%%%%
%%%%%%%%%%%%%%%%%%%%%%%%%%%%%%%%%%%%%%%%%%%%%%%%%%%%%%%%%%%%%%%%%%%%%%%%
Obviously, the typical non-zero jump lengths can not exceed the width of the critical zone, $\ell_x$, therefore $\kappa\le 2/3$. 
Here we conjecture that the jump lengths are in the order of the correlation length $\ell_x$, and thus 
\be 
\kappa=\alpha.
\label{kappaalpha}
\ee 

\begin{figure}[ht!]
\includegraphics[width=8.5cm]{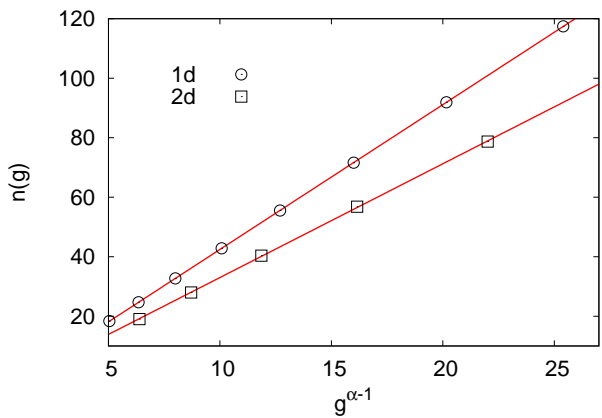} 
%jumpnum.eps
\caption{\label{jumpnum} The average number of non-zero jumps of the front under a change $\Delta x_c=1/g$, plotted against $g^{\alpha-1}$ for the one-dimensional and the two-dimensional model with the corresponding value of $\alpha$. For the two-dimensional model, jumps of the local front coordinates are considered. The straight lines are linear fits to the data. 
}
\end{figure}

So far we considered an elementary, unit shift in $x_c$, corresponding to a change of $g$ in the order parameter $\overline{\Delta_x}$, see Eq. (\ref{deltax}). 
If the order parameter $\overline{\Delta_x}$ is changed by $O(1)$, the front moves forward by $\Delta x_f\sim g^{-1}$ and, on average, this $O(g^{-1})$ front displacement is realized in distinct jumps,
\be
n(g)\sim {\rm Prob(l>0)}/g\sim g^{\alpha-1}
\label{ng}
\ee 
in number. 
This is in accordance with numerical results shown in Fig. \ref{jumpnum}, where the average number of non-zero jumps of the front, ${\rm Prob(l>0)}/g$, during a change $\Delta x_c=1/g$ is plotted against $g^{\alpha-1}$.
%%%%%%%%%%%%%%%%%%%%%%%%%%%%%%%%%%%%%%%%%%%%%%%%%%%%%%%%%%%%%%%%%%%%%%%%%
%%%%%%%%%%%%%%%%%%%%%%%%%%%%%%%%%%%%%%%%%%%%%%%%%%%%%%%%%%%%%%%%%%%%%%%%

\section{Disordered contact process in two dimensions}
\label{sec:2d}

\subsection{Width of the front}

In this section, we consider the more realistic, two-dimensional, DCP in the presence of a gradient. As the colonized cluster is generally not connected, it is not possible to delineate it by a single connected boundary. Of course, one could still consider the percolation hull of the colonized sites, in the same way as for the clean model. However, by this definition, some ``outposts'' which are physically part of the CC but not connected with it in the percolation sense, would fall in the outer side of the hull. It is thus an improper choice for the characterization of the disordered problem. Here, we study the set of $x$ coordinates of outermost sites for each value of the coordinate $y$.  

\begin{figure}[ht!]
\includegraphics[width=8.5cm]{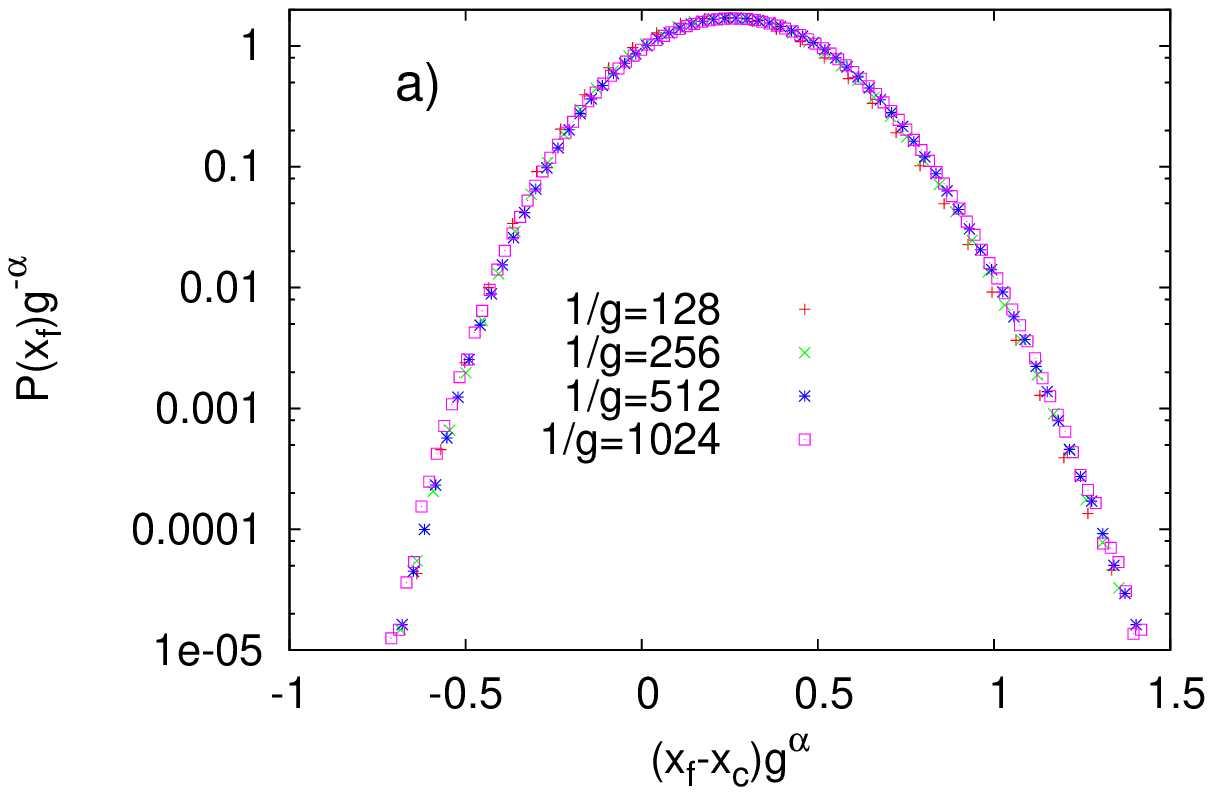} 
\includegraphics[width=8.5cm]{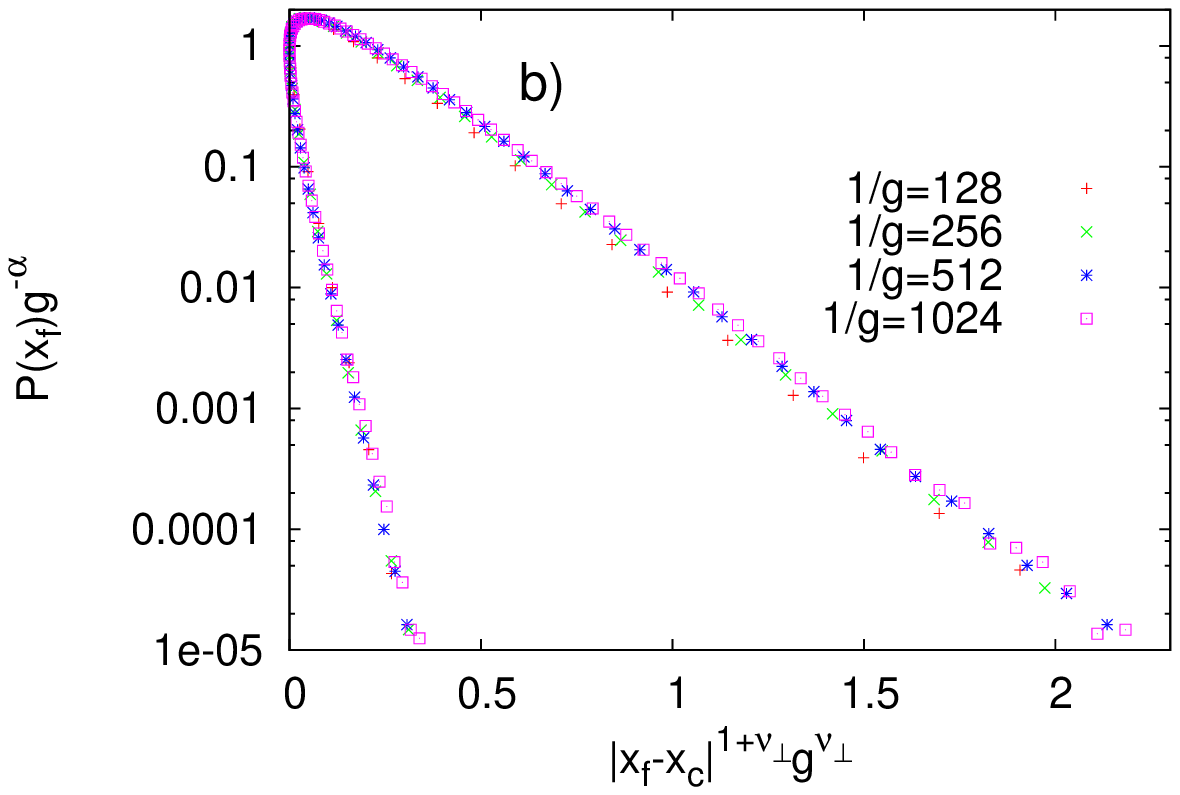} 
%xfd.eps
%xfd2.eps
\caption{\label{2ddist} a) The distribution of the $x$ coordinate of the front, $x_f$ in the two-dimensional DCP, obtained by the SDRG method. The number of samples was $8\cdot 10^4$ for each value of the gradient. The exponents $\nu_{\perp}=1.24$, $\alpha=0.554$ are used, see Table \ref{table_exponents}. 
b) The same data plotted against $|x_f-x_c|^{1+\nu_{\perp}}g^{\nu_{\perp}}$. In this plot, the tail of the distribution must be linear according to Eq. (\ref{Pl}).
}
\end{figure}

We performed numerical SDRG calculations in samples of size $L\times L$, with $g=1/L$, and collected the front coordinates $x_f$ at all transversal coordinate $y$. 
The distributions are shown in Fig. \ref{2ddist}. 
%%%%%%%%%%%%%%%%%%%%%%%%%%%%%%%%%%%%%%%%%%%%%%%%%%%%%%%%%%%%%%%%%%%%%%%%%
%%%%%%%%%%%%%%%%%%%%%%%%%%%%%%%%%%%%%%%%%%%%%%%%%%%%%%%%%%%%%%%%%%%%%%%%
As can be seen, using the correlation-length exponent of the two-dimensional DCP quoted in Table \ref{table_exponents}, the distributions follow the general scaling law in Eq. (\ref{ellg}). Furthermore, the tails of the distribution are compatible with the compressed exponential form in Eq. (\ref{Pl}).
In contrast to the one-dimensional model, the distributions are not symmetric around $x_c$ and the most likely position is in the active domain ($x>x_c$). 

In a system of finite transversal size $L_y$, one can also consider the $x$ coordinate of the outermost site of the CC in the whole sample, $x_g=\min_{1\le y\le L_y}\{x_f(y)\}$, corresponding to the global boundary. Then, the approximate distribution of $x_g$ can be obtained by extreme-value analysis. 
For this, we have to take into account that the front coordinates $\{x_f(y)\}$ in a given sample are correlated on a scale $\ell_y$, see Eq. (\ref{elly}).
The effectively independent number of the data is therefore $n_y\sim L_y/\ell_y$. 
Starting with a compressed exponential parent distribution of $l=|x_f-x_c|$, 
\be
P_>(l)=e^{-Cg^{\nu_{\perp}}l^{1+\nu_{\perp}}},
\label{parent}
\ee
it is straightforward to show that the maximal, global deviation $l_g=\max_{1\le n\le n_y}\{l_n\}$ follows,
for  $Cg^{\nu_{\perp}}l_g^{1+\nu_{\perp}}\gg 1$, the Gumbel distribution
\be 
\rho(z)\approx e^{-z-e^{-z}},
\ee
given in terms of the reduced variable $z=Cg^{\nu_{\perp}}l_g^{1+\nu_{\perp}} - \ln n_y$. As $z$ is an $O(1)$ random variable, for $\ln n_y\gg 1$ we obtain that the typical value of the global deviation scales as 
\be 
l_g\sim g^{-\alpha}(\ln n_y)^{1-\alpha}
\sim g^{-\alpha}|\ln g|^{1-\alpha}.
\ee
Here, the second relation is valid for a sample of size $L\times L$, for which $n_y\sim g^{\alpha-1}$, see Eq. (\ref{elly}).

\subsection{Evolution of the front}

\begin{figure}[ht!]
\includegraphics[width=8cm]{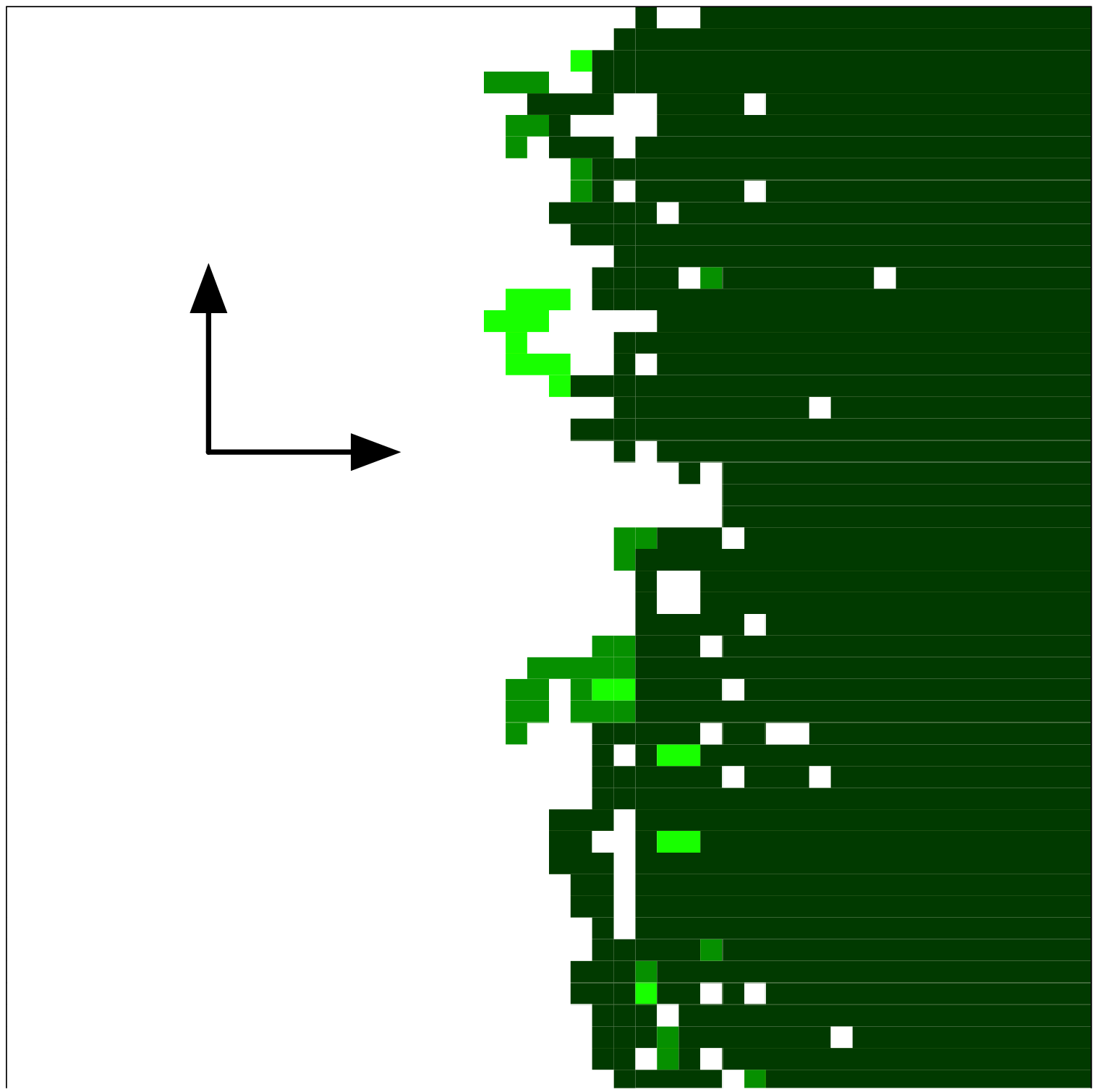}
\includegraphics[width=8cm]{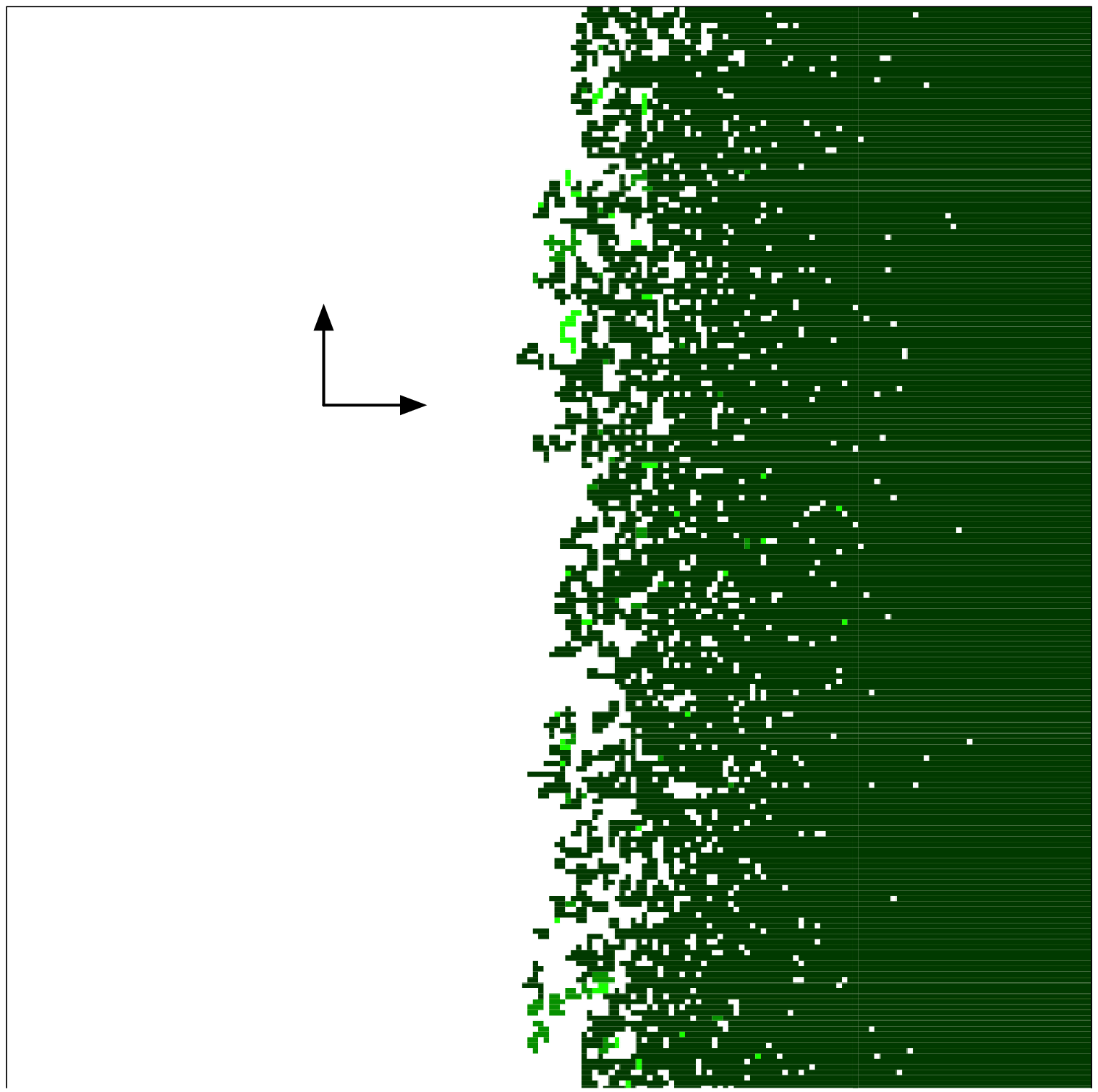}
%ki_50_green.eps
%ki_200_green.eps
\caption{\label{2dchange}
Colonized clusters determined by the SDRG method at three subsequent steps of $x_c$ for $L=50$ (top) and for $L=200$ (bottom). The sites vanishing from the CC at earlier steps (smaller $x_c$) are colored by lighter shades of green. The arrows are proportional to the correlation length.
}
\end{figure}

\begin{figure}[ht!]
\includegraphics[width=8.5cm]{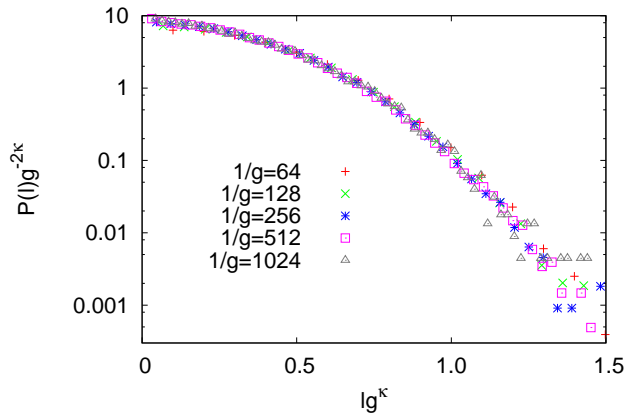} 
%jump.eps
\caption{\label{jump2d} Scaled distributions of the local shift of the front positions for different values of the gradient, obtained for the two-dimensional DCP by the numerical SDRG method. The value of the exponent $\kappa=\alpha=0.554$, see Table \ref{table_exponents}, is used.
}
\end{figure}

We now turn to the question of how the front advances under a global change of the environment. We follow the same protocol as for the one-dimensional model, i.e. the critical coordinate $x_c$ is changed in unit steps, and the set of local front coordinates $\{x_f(y)\}$ is determined by the SDRG method for each $x_c$.
Typical colonized clusters at subsequent steps of $x_c$ are shown in Fig. \ref{2dchange}. 
%%%%%%%%%%%%%%%%%%%%%%%%%%%%%%%%%%%%%%%%%%%%%%%%%%%%%%%%%%%%%%%%%%%%%%%%%
%%%%%%%%%%%%%%%%%%%%%%%%%%%%%%%%%%%%%%%%%%%%%%%%%%%%%%%%%%%%%%%%%%%%%%%%
We monitored the local shift $l$ of $x_f(y)$ for fixed transversal coordinates $y$, as well as the global shift $l_g$ of the extremal coordinate $x_g$. Note that in the latter case the $y$ coordinate of the globally outermost site is not fixed.
The local shift distributions calculated for each $y$ are shown in Fig. \ref{jump2d} in a few thousands of samples for each $g$.  
%%%%%%%%%%%%%%%%%%%%%%%%%%%%%%%%%%%%%%%%%%%%%%%%%%%%%%%%%%%%%%%%%%%%%%%%%
%%%%%%%%%%%%%%%%%%%%%%%%%%%%%%%%%%%%%%%%%%%%%%%%%%%%%%%%%%%%%%%%%%%%%%%%
Similarly to the one-dimensional model, the local shifts follow the intermittent dynamics described by Eqs. (\ref{prob0}-\ref{Plg}), and a good scaling collapse of the distributions is achieved by  $\kappa=0.554$, in accordance with the generalization of our conjecture $\kappa=\alpha$ formulated in one dimension.
We can also study the number of jumps to achieve a total displacement $\Delta x_f\sim g^{-1}$ of the local front position, corresponding to an $O(1)$ change in the control paremeter. Numerical results for the average number of non-zero jumps of the local front coordinate shown in Fig. \ref{jumpnum} are in accordance with the expectation $O(g^{\alpha-1})$, see Eq. (\ref{ng}). 

\begin{figure}[ht!]
\includegraphics[width=8.5cm]{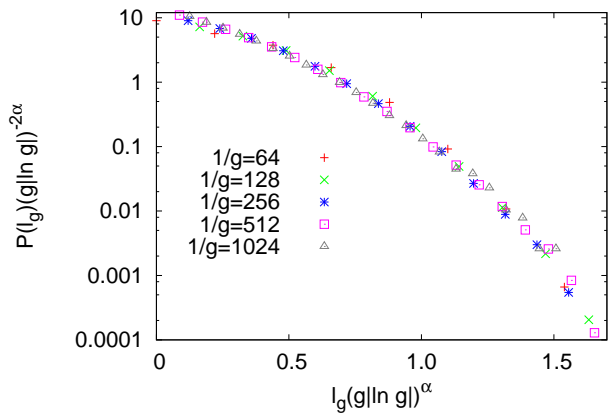} 
%jumpg.eps
\caption{\label{jumpg2d} Scaled distributions of the global shifts for different values of the gradient, obtained for the two-dimensional DCP by the numerical SDRG method. The value of the exponent $\kappa=\alpha=0.554$, see Table \ref{table_exponents}, is used.
}
\end{figure}

\begin{figure}[ht!]
\includegraphics[width=8.5cm]{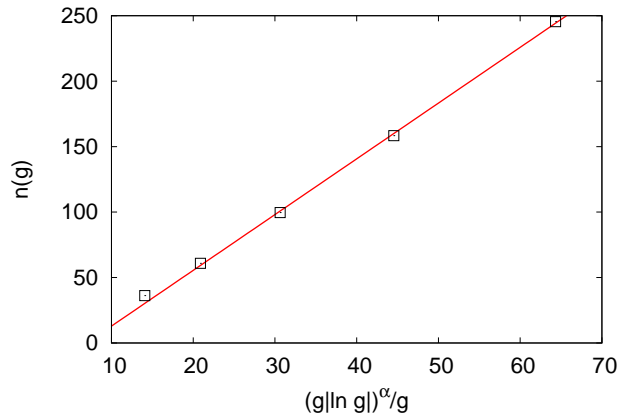}
%jumpnumg.eps 
\caption{\label{jumpnumg} The average number of non-zero global jumps of the front under a change $\Delta x_c=1/g$, plotted against $(g|\ln g|)^{\alpha}/g$ for the two-dimensional model. The straight line is a linear fit to the data. 
}
\end{figure}

Concerning the global jumps $l_g$, we make the following considerations. 
When the outermost site disappears (together with a bunch of sites in the correlation volume $\ell_x\times\ell_y$) under the change of $x_c$, the next outermost site will be typically in a different correlation volume of the front region. The length $l_g$ of a global jump (once it occurs) is thus typically given by the difference of the smallest and the second smallest front coordinate $x_f(y)$ in the sample. Assuming again independent variables, $n_y$ in number, one obtains in straightforward calculations that the distribution of the ``gap'' $l_g$ between the first and second extrema has the distribution
\be 
p_{n_y}(l_g)=n_y(n_y-1)\int_0^{\infty}[P_<(x)]^{n_y-2}\rho(x+l_g)\rho(x)dx,
\ee 
where $P_<(x)=1-P_>(x)$ is the parent distribution given in Eq. (\ref{parent}) and $\rho(x)$ is the corresponding probability density. Here, we assumed for the sake of simplicity that the parent distribution is continuous.   
Although the integral cannot be evaluated with the distribution in Eq. (\ref{parent}), the scaling of $l_g$ can be determined by the saddle-point approximation. It is straightforward to show that, under the condition $\ln n_y\gg 1$, the saddle point of the integrand is at 
$x^*\approx \left[\frac{g^{-\nu_{\perp}}}{C}\ln(n_y/2)\right]^{1-\alpha}$,
and the tail of the distribution behaves as
\be 
p_{n_y}(l_g)\sim \exp\{-C'[g\ln(n_y/2)]^{\alpha}l_g\},
\ee
up to exponential precision. 
In square samples, for which $n_y\sim g^{\alpha-1}$ we thus expect the rare, non-zero global jumps to scale as 
\be 
l_g\sim (g|\ln g|)^{-\alpha}.
\ee
Accordingly, the global shift $l_g$ obeys similar scaling laws as in Eqs. (\ref{prob0}-\ref{Plg}) with the difference that $g$ is replaced by $g|\ln g|$. 
This is supported by the numerical distributions shown in Fig. \ref{jumpg2d}, with a good scaling collapse.  
%%%%%%%%%%%%%%%%%%%%%%%%%%%%%%%%%%%%%%%%%%%%%%%%%%%%%%%%%%%%%%%%%%%%%%%%%
%%%%%%%%%%%%%%%%%%%%%%%%%%%%%%%%%%%%%%%%%%%%%%%%%%%%%%%%%%%%%%%%%%%%%%%%
The average number of global jumps under an $O(1)$ change of the global control parameter is expected to scale as 
$n(g)\sim {\rm Prob(l>0)}/g\sim (g|\ln g|)^{\alpha}/g$, 
in agreement with numerical data presented in Fig. \ref{jumpnumg}.
%%%%%%%%%%%%%%%%%%%%%%%%%%%%%%%%%%%%%%%%%%%%%%%%%%%%%%%%%%%%%%%%%%%%%%%%%
%%%%%%%%%%%%%%%%%%%%%%%%%%%%%%%%%%%%%%%%%%%%%%%%%%%%%%%%%%%%%%%%%%%%%%%%

\section{Conclusion}
\label{sec:summary}

Motivated by the problem of range margin in an environmental gradient, studied intensively in ecology, we considered in this work the effects of quenched disorder by studying the disordered contact process with a linear spatial trend in the local control parameter.     
We applied the SDRG approach to the model, enabling us to construct the colonized cluster, i.e. the set of sites occupied with a high probability. 
The CC is a disconnected set, and we identified its front as the outermost constituent of the CC. 

In one dimension, we have shown that the problem of finding the front position is equivalent to an extremal problem of a random walk with a time-dependent bias. We confirmed by explicit calculations that the distribution of the front position for small gradients is in agreement with the general scaling considerations, which predict a compressed exponential tail of the distribution, and in which the model-dependence enters through the correlation-length exponent. 
 Concerning the shift of the front under a unit change of the critical coordinate, we demonstrated that the front advances intermittently, meaning that it remains unchanged in almost all steps, while it makes long jumps in rare steps, the fraction of which vanishes with decreasing gradient. 
These characteristics of the intermittent motion of the front are found to follow universal scaling laws, conjectured to be related to the correlation-length exponent of the model. 

In two dimensions, we studied both the coordinates of outermost sites of the CC for a fixed transversal coordinate, as well as its globally extremal value, by applying the SDRG method numerically. The distribution of local front coordinates was found to follow the general scaling law.   
Implementing the same protocol as in one dimension, the corresponding local and global shifts were measured. These are found to show the same intermittent motion as in the one-dimensional model, and to obey the same scaling laws (including a logarithmic correction for the global shift) with the corresponding correlation-length exponent.

It is worth comparing the motion of the boundary under the change of control parameter in two dimensions to the thoroughly studied problem of driven interfaces in random media \cite{fisher,giamarchi}.
The basic ingredients of that model at zero temperature, written as an overdamped dynamical equation, are the elastic interactions between the constituents of the interface, which tend to keep the interface flat, and the static random pinning forces of the medium. As a response to an external force, provided it exceeds a critical depinning force, the interface performs a persistent but jerky motion, of which the intermittent motion of the boundary in our model is reminiscent. 
Beside the similar appearances, there are, however, several differences between the two models. A formal one is that the description of the dynamics of the boundary in our model by an autonomous dynamical equation in terms of the degrees of freedom of the boundary is unresolved. Furthermore, the boundary, as it is defined, is not necessarily a connected object. Finally, the mean velocity of the boundary (or the change rate of the control parameter) is an external parameter in our model, as opposed to driven interfaces where it is a response to the applied force.

It would be interesting to examine whether quenched disorder has observable effects in real ecological systems. As the difference between the characteristics of clean and quenched disordered systems is most prominent at criticality, an ideal testing ground for disorder effects would be provided by ecological systems subject to an environmental gradient, since here a critical region naturally appears without any fine-tuning of a control parameter. 
In particular, it would be interesting to validate our key result, the intermittent evolution of the front under global climate change by empirical data.
Our results indicate that climate change effects might stay concealed for an extended period of time, leading to sudden large changes in the observations.

%%%%%%%%%%%%%%%%%%%%%%%%%%%%%%%%%%%%%%%%%%%%%%%%%%%%%%%%%%%%%%%%%%%%%%%%%%%%
%%%%%%%%%%%%%%%%%%%%%%%%%%%%%%%%%%%%%%%%%%%%%%%%%%%%%%%%%%%%%%%%%%%%%%%%%%%%

\begin{acknowledgments}
We thank B. Oborny, M. Gastner, G. \'Odor, and F. Igl\'oi for useful discussions. This work was supported by the National Research, Development and Innovation Office NKFIH under grant No. K128989. I.A.K. was supported by the Domus Hungary Scholarship of the Hungarian Academy of Sciences.
This publication was made possible through the support of a grant from the John Templeton Foundation. The opinions expressed in this publication are those of the authors and do not necessarily reflect the views of the John Templeton Foundation.
\end{acknowledgments}

\appendix
\section{Distribution of the front}
\label{app:distribution}

Let us first assume that the local control parameter is discontinuous at $x_c$, while constant otherwise: $\overline{\Delta}(x)=\Delta_1<\Delta_c$ for $x<x_c$ and $\overline{\Delta}(x)=\Delta_2>\Delta_c$ for $x>x_c$. 
The front is then pinned on average to $x_c$ and its penetration into either phase in a distance $l=|x-x_c|$ is exponentially improbable: 
\be
P(l)\sim e^{-l/\xi_i},
\ee
where $\xi_i$ is the correlation length $\xi_i\sim |\Delta_i-\Delta_c|^{-\nu_{\perp}}$, for $i=1,2$. 
In a medium with a continuously varying control parameter this can be generalized to 
\be 
\ln P(l) \sim -\int^ll'/\xi(l')dl', 
\ee
which, substituting $\xi(l)\sim [\Delta(l)-\Delta_c]^{-\nu_{\perp}}\sim (gl)^{-\nu_{\perp}}$, 
results in a compressed exponential tail of the distribution given in
Eq. (\ref{Pl}). 
This result can also be confirmed by formulating lower and upper bounds on $P(l)$ as will be done for a similar function in section \ref{app:rw}.

\section{The front position in one dimension and time-dependent random walks}
\label{app:rw}

The proof of the statement formulated in section \ref{subsec:1dwidth} is simple. Clearly, the maximum point must be at an even index, since the terms $\beta_n$ are alternately positive and negative for even and odd $n$, respectively. 
Assume that the maximum point is at $2m$. It is obvious from Fig. \ref{zigzag}a, that the SDRG procedure always just eliminates some $Y_n$, so, until $Y_{2m}$ is not removed it will remain the maximal value at any stage of the procedure. 
But its removal can never happen, which can be seen as follows.
The extinction rate of site $m+1$ is represented by the descending segment of path between $2m$ and $2m+1$, see Fig. \ref{zigzag}b. The removal of site $m+1$ could only happen if both of its neighboring ascending segments were higher in magnitude. This, however, cannot be fulfilled for its right neighbor since then $Y_{2m+2}$ would be greater than $Y_{2m}$, contradicting our assumption.
Therefore site $m+1$ is never removed during the SDRG procedure.
A fusion of site $m+1$ (or the cluster containing site $m+1$) with a cluster on its left would require that the height of the ascending segment on the left of the maximum be smaller in magnitude than those of its neighboring descending segments. However, this cannot be true for its left descending neighbor since then $Y_{2m-2}$ would be greater than $Y_{2m}$.  This is again in contradiction with our assumption. Consequently no sites with $n<m+1$ are fused to site $m+1$. We thus conclude that site $m+1$ is the outermost site of the colonized cluster, i.e. $x_f=m+1$.  

In the sequel, we will make two simplifications. First, as we are interested in the scaling of the front position for small $g$, where $x_f$ is typically large, we can write $x_f\approx m$. Second, we will consider the series $Y_n$ restricted to even indices, $n=2x$, and replace the indices $2x$ with $x$. 
The problem of determining $x_f$ can thus be summarized as follows. We have a random walk with a time-dependent bias given in Eq. (\ref{bias}) which changes sign from positive to negative at time $x=0$, and we are looking for the time $x_f$ at which the walker reaches the farthest position in the positive direction. 

\begin{figure}[ht]
\begin{center}
\includegraphics[width=8.5cm]{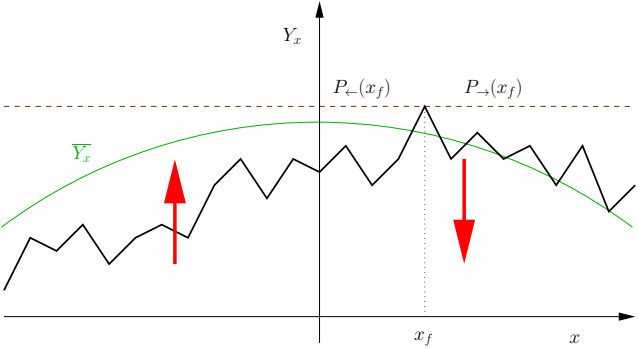} 
%px2.eps
\caption{\label{px} Illustration of a typical path $Y_x$ of a random walk (black line), which is biased in the positive (negative) direction for times $x<0$ ($x>0$). The direction of the bias is indicated by the red arrows. 
The dashed horizontal line represents the absorbing wall for the survival problem, see the text.
The solid green curve is the average path given in Eq. (\ref{Yav}).   
}
\end{center}
\end{figure}

Clearly, as the bias is symmetric in time, the distribution of $x_f$, $P(x_f)$ 
is an even function, i.e. 
\be
P(x_f)=P(-x_f), 
\label{symm}
\ee
and the mean value $\overline{x_f}=0$. Furthermore, the distribution $P(x_f)$ is expected to have a maximum   
at $x_f=0$, since, for $x_f\neq 0$, the walker moves against the bias during the time interval $|x_f|$.  
Due to the symmetry of $P(x_f)$, it is enough to deal with the positive part $x_f>0$ in the following. 
Now observe that the probability that the maximum is reached at time $x_f$, is equivalent with the probability that the path $Y_x$ remains below $Y_{x_f}$ in both time directions starting from $x_f$, see Fig. \ref{px}. 
%%%%%%%%%%%%%%%%%%%%%%%%%%%%%%%%%%%%%%%%%%%%%%%%%%%%%%%%%%%%%%%%%%%%%%%%%
%%%%%%%%%%%%%%%%%%%%%%%%%%%%%%%%%%%%%%%%%%%%%%%%%%%%%%%%%%%%%%%%%%%%%%%%
That means, $Y_{x_f\pm x}<Y_{x_f}$ for all $x=1,2,\cdots$. Thus $P(x_f)$ is a product of two survival probabilities:
\be 
P(x_f)=P_{\leftarrow}(x_f)P_{\rightarrow}(x_f)
\ee
Here, $P_{\rightarrow}(x_f)$ is the survival probability of a random walk in a time-dependent bias starting at time $x_f$, whereas $P_{\leftarrow}(x_f)$ is the survival probability of a random walk starting at time $x_f$ and moving in reversed time, toward $x=0$ and beyond. 
The second factor, $P_{\rightarrow}(x_f)$, describes a situation where the walker is biased away from the absorbing wall it has to survive. 
For the first factor, $P_{\leftarrow}(x_f)$, the walker is initially attracted toward the wall from time $x_f$ to $0$ (keep in mind that time is reversed), and subsequently, for times $x<0$ it is repelled away from the wall.
As the more rapidly varying factor is the second one, we assume that the scaling properties of $P(x_f)$ are determined by those of $P_{\leftarrow}(x_f)$ up to an exponential precision.
$P_{\leftarrow}(x_f)$ can be decomposed as $P_{\leftarrow}(x_f)=P_+(x_f)P_-(x_f)$, where $P_+(x_f)$ is the probability that the walk starting at time $x_f$ survives up to time $0$, whereas $P_-(x_f)$ is the conditional probability that under the survival down to time $0$, the walk 
will survive throughout in the regime $x<0$. Clearly, $P_-(x_f)$ will also depend on $x_f$ through the position of the walker at time $x=0$ which depends on $x_f$. But, as the walker arrives at time $x=0$ through a time domain in which it is attracted toward the wall, the above dependence is expected to be weak, and remaining within the exponential precision, we may write 
\be 
P_{\leftarrow}(x_f)\sim P_+(x_f). 
\ee
The calculation of $P_+(x_f)$, which is the survival probability of a time-dependent random walk is still a hard problem for which, up to our knowledge, no general results exist. Nevertheless, we can formulate lower and upper bounds of $P_+(x_f)$ and will see that both have the same scaling property. 

This can be done as follows. $P_+(x_f)$ describes the survival probability in a time-dependent bias given by the series $gx_f,g(x_f-1),\dots,0$. 
A lower bound is provided by the survival probability in $x_f$ steps in the presence of a constant bias $gx_f$, $P_s(\Delta=gx_f,t=x_f)$. An upper bound can be obtained by a survival probability in a shorter time interval $x_f/2$ in the presence of a smaller constant bias $gx_f/2$, $P_s(\Delta=gx_f/2,t=x_f/2)$. 
So we have 
\be 
P_s(\Delta=gx_f,t=x_f)<P_+(x_f)<P_s(\Delta=gx_f/2,t=x_f/2).
\ee
Now we make use of the result that in the presence of a constant bias toward the absorbing wall, the survival probability decays exponentially in time 
\be 
P_s(\Delta,t)\sim e^{-t/\tau}, 
\ee
where the correlation time diverges according to
\be 
\tau\sim\Delta^{-2}
\label{tau}
\ee
as $\Delta\to 0$. This result was derived in Ref. \cite{ri} for random walks with a particular (bimodal) distribution of step lengths but, due to universality, it is expected to hold generally. 
We have thus for the lower bound $P_s(\Delta=gx_f,t=x_f)\sim e^{-Cg^2x_f^3}$, and the upper bound $P_s(\Delta=gx_f/2,t=x_f/2)\sim e^{-C'g^2x_f^3}$, where $C$ and $C'=C/8$ are constants. As a consequence, $P_+(x_f)$, as well as $P(x_f)$ must have the scaling form given in Eq. (\ref{Pxf}).

\section{Validity of the quasistatic approximation}
\label{app:qs}

In this section, we derive a condition for the change rate of the control parameter, $v_{\Delta}=\frac{d\overline{\Delta_x}(t)}{dt}$, valid for both one and two dimensions, under which the process can be regarded as quasistatic. 
We have seen that the jumps of the front are rare; if the critical coordinate advances by $\ell_x\sim g^{-\alpha}$, only one jump [of length $O(\ell_x)$] occurs typically. This event is realized by the vanishing of the cluster extending from the old to the new position of front, having a linear size $O(\ell_x)$. This cluster lies in the critical zone around $x_c$, therefore the time scale of this event is well approximated by the lifetime $\tau$ of the critical DCP in a finite sample of linear size $\ell_x$. This is known to scale as $\ln\tau\sim\ell_x^{\Psi}$, where the barrier exponent $\Psi$ is $1/2$ in one dimension \cite{hiv} and close to $1/2$ in two dimensions \cite{kovacs}.   
A quasistatic change in a weak sense is then achieved if the typical time between subsequent jumps of the front, $t\sim g\ell_x/|v_{\Delta}|$, is much longer than the relaxation time: $t\gg \tau$. This leads to the condition for the change rate of the control parameter
\be 
|v_{\Delta}|\ll g^{1-\alpha}e^{-Cg^{-\alpha\Psi}}.
\ee   
If this condition is fulfilled, the system has sufficient time to relax before the next jump of the front occurs, although, there will be a time lag of $O(\tau)$ in the evolution of the front position, $x_f(t)$, compared to the instantaneous stationary value. 
A more stringent condition of a quasistatic change is obtained by the requirement that the time $t$ under which the critical coordinate makes a unit displacement, $\Delta x_c=1$, is much longer than the relaxation time. This imposes the following condition for the change rate of the control parameter: 
\be
|v_{\Delta}|\ll ge^{-Cg^{-\alpha\Psi}}.
\ee

%%%%%%%%%%%%%%%%%%%%%%%%%%%%%%%%%%%%%%%%%%%%%%%%%%%%%%%%%%%%%%%%%%%%%%%%%%%%%%
%%%%%%%%%%%%%%%%%%%%%%%%%%%%%%%%%%%%%%%%%%%%%%%%%%%%%%%%%%%%%%%%%%%%%%%%%%%%%%

\end{document}